\documentstyle[preprint,aps]{revtex}
\begin{document}
\draft
\preprint{TMU-NT970606}
%
%
%
%
%
\title{Dissipative Field Theory 
                 with Caldeira-Leggett Method
       and its Application 
                 to Disoriented Chiral Condensation}
\author{Hiroyuki Yabu, Satoshi Nozawa and 
        Toru Suzuki\thanks{Supported by ME grant.}}
\address{Department of Physics, 
         Tokyo Metropolitan University, 
         Hachioji Tokyo 192, Japan}
\date{\today}
\maketitle
\begin{abstract}
     The effective field theory including the dissipative effect 
is developed based on the Caldeira-Leggett theory 
at the classical level. 
After the integration of the small field fluctuations 
considered as the field radiation, 
the integro-differential field equation is given 
and shown to include the dissipative effects. 
In that derivation, special cares should be taken 
for the boundary condition of the integration. 
Application to the linear sigma model 
is given, 
and the decay process of the chiral condensate 
is calculated with it, 
both analytically in the linear approximation 
and numerically. 
With these results, 
we discuss the stability of chiral condensates 
within the quenched approximation.
\end{abstract}
%
%
%
%
%
%
\section{Introduction}
%
%
     Recently, there has been great interests 
in the collective phenomena 
that might be observed after the heavy ion collisions, 
especially in the state called 
the disoriented chiral condensate (DCC) \cite{BKT,RW}. 

     The quantum chromodynamics (QCD) 
that describes the strong-interaction physics 
has the chiral symmetry, $SU_L(2) \otimes SU_R(2)$, 
that is explicitly broken with the order of the pion mass $m_\pi$. 
The behavior of the vacuum state can be represented 
by the order-parameters 
${\langle \sigma \rangle} ={\langle \bar{q} q \rangle}$ 
and 
${\langle \pi \rangle} ={\langle \bar{q} \gamma_5 q \rangle}$ 
where $q$ is a quark field.   
At zero-temperature, 
the effective potential 
$V_{\rm eff}({\langle \sigma \rangle},{\langle \pi \rangle})$
has a bottom circle with the radius $f_\pi$, 
so that this symmetry is spontaneously broken 
and the vacuum state takes 
${\langle \sigma \rangle} \neq 0$ and 
${\langle \pi \rangle}=0$.  

     The DCC state is defined as that
on the bottom circle of $V_{\rm eff}$ but 
with ${\langle \pi \rangle} \neq 0$,  
and expected to be produced after the high-energy proton 
or heavy-ion collision. 
About the formation of the DCC, there has been a lot of discussions, 
especially concerning with the quenching or annealing scenarios 
\cite{RW,GGP,GM,GGM,AHW}, 
but the present situation is still controversial. 

     As has been pointed out in \cite{GVN}, 
another important problem is the decay process of the DCC state. 
Because of the explicit breaking of the chiral symmetry, 
the DCC state is only meta-stable and considered to decay 
into the true vacuum 
with radiating many number of coherent pions \cite{AR}. 
If the life-time of the DCC is very short 
compared with the formation time, 
it will be very difficult to observe the DCC state 
through the pion signal. 

     The entire process of the DCC 
can be formulated as the dissipative system 
of the collective coordinates 
(the order parameters 
${\langle \sigma \rangle}$ and 
${\langle \pi \rangle}$) 
under the back-ground pion radiation, 
so that we can analyze the DCC, 
especially the decay of it, with the method of the dissipative theory 
in the semiclassical dynamics. 

     There has been a long history to derive the equation of motion 
with the radiative damping in the classical electrodynamics 
and the thermo-statistical theory. 
The motion of the charged particle with the electromagnetic radiation 
had been much discussed since pre-quantum mechanical era \cite{PL}. 
Here we should mention that the advanced Green function 
for the radiation was introduced by Dirac \cite{DR}
to cancel the divergent electron's radiation mass 
and was justified by Wheeler and Feynman \cite{WF}. 

     In the thermo-statistical theory, the dissipative dynamics 
has been considered with the Brownian motion 
and the Langevin equation has been one of the most important tool 
to attack the problem \cite{WE}:
\begin{equation}
     M \ddot{q} +\gamma \dot{q} +V'(q) =R(t),  
\label{EqInt}
\end{equation}
where $q$ and $R$ are the particle coordinate 
and the fluctuating force 
and $\gamma$ is the dissipative coefficient. 
However, in the Langevin theory, the dissipative coefficient 
is the given parameter and can not be fixed by itself. 

     In 1983, Caldeira and Leggett proposed a method to derive 
the dissipation from the microscopic theory \cite{CL}. 
They start with the system-plus-reservoir model 
(the interacting system of the collective coordinates 
and the background freedom) 
and, with integrating out the background freedom, 
the dissipative equation for the collective coordinates 
can be obtained. 
This method was formulated in the particle dynamics, 
and has been applied for the polaron motion in a crystal 
(acoustic polaron), 
diffusion of charged interstitials in normally conducting metals 
and the dynamics of Josephson junction \cite{WE}. 
In the latter case, 
the reduction of the tunneling probability was predicted 
for the macroscopic quantum tunneling effect 
between superconductor and insulator 
with the Caldeira-Leggett theory
and it was really confirmed experimentally \cite{VW,MDC}.

     The application of the Caldeira-Leggett theory 
to the nuclear/hadron physics 
is very interesting: especially to the phenomena 
with the multi-particle 
production such as the emission of meson, photon and so on. 
The energy of the system 
(the excited nucleus or the fireballs, for example) 
is considered to decrease `'dissipatively'' 
with the particle emission 
and the Caldeira-Leggett theory can be applied to them,  
where the reservoir corresponds to the emitted particle. 
Another interesting application is for the DCC phenomenon. 
The DCC process is described by the non-linear sigma model, 
and the collective coordinates and the reservoir are 
just the condensates 
${\langle \sigma \rangle}$ and 
${\langle \pi \rangle}$, 
and the radiative pion through the formation and decay processes.

     In those phenomena, the dynamical variables of the system 
are the fields corresponding to the emitted particles. 
For the application to such a process, 
we have to extend the Caldeira-Leggett theory for the field dynamics. 
That extension is very interesting by itself 
because it produces the dissipative field equation. 

     In this paper, we formulate the Caldeira-Leggett theory 
in the field dynamics, and apply it to the DCC process. 
The extension of the theory for the field theory will be given 
in the first half of this paper as generally as possible.  
The resultant formulation is applied for the non-linear sigma model, 
and the existence of the dissipative effects is shown 
analytically and numerically. 
In the final part of the paper, 
the decay process of the DCC is discussed qualitatively 
within the present formulation.
%
%
\section{Dissipative Field Equation with Caldeira-Leggett Theory}
%
%
     To illustrate the application of the Caldeira-Leggett theory 
to the field equation, 
we take a simple system of the order-parameter $\Phi(x)$ 
and the background field $f(x)$ that describes 
the particle with the mass $m$:
\begin{equation}
     L ={1 \over 2} (\partial\Phi)^2 -V(\Phi) 
       +G(\Phi) f 
       +{1 \over 2} [(\partial{f})^2-m^2 f^2]. 
\label{eQAa}
\end{equation}
where $V(\Phi)$ is the effective potential for $\Phi$ 
and $G[\Phi] f$ gives the interaction 
between $\Phi$ and $f$. 
Taking variations with $\Phi$ and $f$, 
we obtain the field equations from (\ref{eQAa}):
\begin{mathletters}
\label{eQAbc}
\begin{equation}
     \partial^2\Phi 
          +{\delta{V} \over \delta\Phi} 
          +{\delta{G} \over \delta\Phi} f =0, 
\label{eQAb}
\end{equation}
\begin{equation}
     (\partial^2+m^2) f +G[\Phi] =0.      
\label{eQAc}
\end{equation}
\end{mathletters}
They  are easily checked to satisfy 
the energy-momentum conservation. 

     Eq. (\ref{eQAc}) is formally solved 
with the Green functions $G(x-y;m)$ that satisfies 
\begin{equation}
     (\partial^2+m^2) G(x-y) =-\delta^4(x-y).  
\label{eQAd}
\end{equation}
The fluctuation $f$ is then written by\footnote{
     The homogeneous part can be dropped safely 
     because it can be absorbed 
     with proper adjustment of the boundary condition.}
\begin{equation}
     f(x) =\int{d^4y} G(x-y;m) G[\Phi(y)],  
\label{eQAe}
\end{equation}
and, substituting it into (\ref{eQAb}), 
we obtain the integro-differential equation for $\Phi$
\begin{equation}
     \partial^2\Phi(x)
          +{\delta{V[\Phi(x)]} \over \delta\Phi} 
          +{\delta{G[\Phi(x)]} \over \delta\Phi} 
           \int{d^4y} G(x-y;m) G(\Phi[y]) =0.  
\label{eQAf}
\end{equation}
This equation is essentially equivalent 
with (\ref{eQAb}) and (\ref{eQAc}). 

     The Green function in (\ref{eQAf}) can be written by
\begin{equation}
     G(x-y;m) =\int{d^4k \over (2\pi)^4} 
               G(k;m) e^{-ik(x-y)},  
\label{eQAg}
\end{equation}
where $G(k;m)$ is the Fourier transform of $G(x-y;m)$ 
and generally takes complex value. 
In the Caldeira-Leggett theory, 
the imaginary component of $G(k;m)$ can be considered 
as the dissipative term 
caused by the emission of particles for the background field $f$. 
It is suggested by the analysis of the simple Newtonian equation 
including the dissipative terms: 
$m \ddot{x} +\eta \dot{x} =F$; 
After Fourier transformation, 
the dissipative term $\eta \dot{x}(t)$ gives 
that of pure imaginary $i \eta \omega x(\omega)$ 
with the spectral function $x(\omega)$ for $x(t)$. 
Further, we can interpret this dissipation 
as the particle emission 
because ${\rm Im}{G(k;m)}$ includes $\delta^4(k^2-m^2)$ 
(on-mass shell). 
The real part ${\rm Re}{G(k;m)}$ is considered to give 
the modification of the effective potential 
that comes from the background absorption and emission, 
and can be absorbed into $V(\Phi)$ by re-definition. 
Thus we can drop it because the parameters in $V(\Phi)$ 
is adjusted from the experimental data. 
Finally, we obtain the dissipative field equation 
corresponding to (\ref{eQAf}):
\begin{equation}
     \partial^2\Phi(x)
          +{\delta{V[\Phi(x)]} \over \delta\Phi} 
          +{\delta{G[\Phi(x)]} \over \delta\Phi} {\tilde f} =0, 
\label{eQAh}
\end{equation}
where
\begin{equation}
     {\tilde f}(x) =\int{d^4y} \int{d^4k \over (2\pi)^4} 
                   i{\rm Im}{G(k;m)} e^{-ik(x-y)}.  
\label{eQAi}
\end{equation}

     Eq. (\ref{eQAi}) is still ambiguous 
because a variety of Green functions 
(advanced, retarded, causal and so on) can be used in it 
and give solutions with different boundary conditions. 
We should select the proper Green functions 
that shows the dissipative effect when $t \to \infty$ 
with real ${\tilde f}$. 
In the application to the linear sigma model, 
it will be shown 
that the advanced Green function satisfies both conditions 
and a proper selection. 
%
%
\section{Application to Linear Sigma Model}
%
%
     We apply the formulation given in the last chapter 
to the linear sigma model by Gel-Mann and L{\'e}vy \cite{GL}
and discuss the dynamical behavior of the order parameter 
of the $SU(2)$ chiral symmetry. 

     The Lagrangian of the sigma model is 
\begin{equation}
     {\cal L} =\frac{1}{2} 
               \left[ (\partial{\sigma})^2 
                     +(\partial{\pi})^2 \right] 
               -V(\sigma,\pi),  
\label{eQBa}
\end{equation}
where $\sigma$ and $\pi=(\pi^1,\pi^2,\pi^3)$ 
are the sigma and the pion fields. 
In this paper, 
we discuss only the phenomena related with the neutral pion 
so that $\pi$ in (\ref{eQBa}) is a single field from now on.  
The application to the charged pion goes in the same way 
but we have to modify the Lagrangian (\ref{eQBa}) 
to include the photon degrees of freedom. 
The effective potential $V(\sigma,\pi)$ is taken 
to the forth order of the fields:
\begin{equation}
     V(\sigma,\pi) =\frac{\lambda}{4} (\sigma^2+\pi^2-\nu)^2 
                -H \sigma,  
\label{eQBb}
\end{equation}
where the parameters $\lambda$, $\nu$ and $H$ are written by
\begin{equation}
     \lambda =\frac{m_\sigma^2 -m_\pi^2}{2f_\pi^2},  \quad
     \nu^2 =f_\pi^2 
            \frac{m_\sigma^2 -3m_\pi^2}{m_\sigma^2-m_\pi^2}, \quad
     H =f_\pi m_\pi^2,  
\label{eQBc}
\end{equation}
with the sigma and pion mass $m_{\sigma,\pi}$ 
and the pion decay constant $f_\pi$.  

     The last term in (\ref{eQBb}) breaks is the chiral symmetry 
explicitly and produce the finite pion mass. 
It is the simplest potential 
that describes the explicit symmetry breaking phenomena, 
and consistent with the low-energy theorem. 

     The fields $\sigma$ and $\pi$ can be divided into 
two parts:
\begin{equation}
     \sigma ={\langle \sigma \rangle} +\delta{\sigma},  \quad
     \pi ={\langle \pi \rangle} +\delta{\pi},  
\label{eQBd}
\end{equation}
where ${\langle \sigma \rangle}$ and 
${\langle \pi \rangle}$ corresponds 
to the order parameter of the chiral symmetry 
for the (disoriented) condensed state $\Psi$: 
${\langle \sigma \rangle} 
={\langle \Psi|{\bar q}q|\Psi \rangle}$ and 
${\langle \pi \rangle}
={\langle \Psi|{\bar q} \tau \gamma_5 q|\Psi \rangle}$ 
with the quark field $q=(u,d)$.
(The condensed states may be described with the coherent or 
the squeezed one as quantum states. \cite{AK})
The $\delta{\sigma}$ and $\delta{\pi}$ are the fluctuations 
around them, 
and represent the sigma and pi meson degrees of freedoms. 
The decomposition (\ref{eQBd}) is essentially the same one 
in Bogoliubov theory for the weak-interacting Bose liquid.  
In that theory, the fluctuation represents elementary excitations 
(phonon/roton) that cause the dissipative effects 
in superfluids \cite{GR}. 

     Minimizing the effective potential $V$, 
we obtain the vacuum state ${|0 \rangle}$ 
with which the order parameters become 
\begin{equation}
     {\langle \sigma \rangle} 
          ={\langle 0|\sigma|0 \rangle} \equiv f_\pi,  \quad
     {\langle \pi \rangle} ={\langle 0|\pi|0 \rangle} =0,  
\label{eQBe}
\end{equation}
and the fluctuations $\delta{\sigma}$ and $\delta{\pi}$
describes the mesons with the mass $m_\sigma$ and $m_\pi$ 
each other. 

     Now we consider the dynamical behaviors of
the disoriented-condensed vacuum state 
${|\Psi \rangle} \neq {|0 \rangle}$ 
(${\langle \sigma \rangle} \neq f_\pi$ and 
 ${\langle \pi \rangle} \neq 0$) 
as the $\Phi$ in the last chapter  
and the fluctuations $\delta{\sigma}$ and $\delta{\pi}$ 
as the background field $f$. 
Substituting Eq. (\ref{eQBd}) into (\ref{eQBa}), 
we obtain 
%
\begin{eqnarray}
     L &=& \frac{1}{2} \left[(\partial{\langle \sigma \rangle})^2 
                            +(\partial{\langle \pi \rangle})^2 \right]
           -V({\langle \sigma \rangle},{\langle \pi \rangle}) 
           +H {\langle \sigma \rangle} \nonumber\\
       &+& \partial{\langle \sigma \rangle} \partial\delta{\sigma} 
          +\partial{\langle \pi \rangle} \partial\delta{\pi} 
\nonumber\\
       &+& \frac{1}{2} \left[ (\partial\delta{\sigma})^2 
                            -m_\sigma^2 \delta{\sigma}^2 \right]
          +\frac{1}{2} \left[(\partial\delta{\pi})^2 
                            -m_\pi^2 \delta{\pi}^2 \right], 
\nonumber\\
\label{eQBf}
\end{eqnarray}
where we assume that the background fields are small 
and describe the real sigma and pi meson degrees of freedom, 
so that we dropped ${\cal O}(\delta{\sigma}^3,\delta{\pi}^3)$-terms 
and the $\delta{\sigma}^2$ and $\delta{\pi}^2$ terms 
are adjusted to be the mass term 
with the real pion and sigma mass $m_\sigma$ and $m_\pi$ 
as in (\ref{eQAa}). 
In this paper, we discuss the case 
where the condensates are 
on the chiral circle: 
${\langle \sigma \rangle}^2+{\langle \pi \rangle}^2 =\nu^2$, 
so that ${\langle \sigma \rangle}$ and ${\langle \pi \rangle}$ 
can be parameterized with the chiral angle $\phi$: 
\begin{equation}
     {\langle \sigma \rangle} =\nu\cos\phi,  \quad
     {\langle \pi \rangle} =\nu\sin\phi.  
\label{eQBg}
\end{equation}
Substituting Eq. (\ref{eQBg}) into (\ref{eQBf}), 
it is obtained that
\begin{eqnarray}
     L &=& \frac{\nu^2}{2} (\partial\phi)^2 +H\nu\cos\phi 
\nonumber\\
       &+& \nu \partial\cos\phi \partial\delta{\sigma} 
          +\nu \partial\sin\phi \partial\delta{\pi} 
\nonumber\\
       &+& \frac{1}{2} \left[(\partial\delta{\sigma})^2 
                           -m_\sigma^2 \delta{\sigma}^2 \right]
          +\frac{1}{2} \left[(\partial\delta{\pi})^2 
                           -m_\pi^2 \delta{\pi}^2 \right]. 
\nonumber\\
\label{eQBh}
\end{eqnarray}

     Taking the variation 
with $\phi$, $\delta{\sigma}$ and $\delta{\pi}$, 
we obtain a set of Euler-Lagrange equations:
\begin{mathletters}
\label{eQBijk}
\begin{equation}
     \nu^2 \partial^2\phi +H\nu\sin\phi 
       -\nu\sin\phi \partial^2\delta{\sigma} 
       +\nu\cos\phi \partial^2\delta{\pi} =0,  
\label{eQBi}
\end{equation}
\begin{equation}
     (\partial^2+m_\sigma^2) \delta{\sigma} 
            +\nu \delta^2\cos\phi =0,      
\label{eQBj}
\end{equation}
\begin{equation}
     (\partial^2+m_\pi^2) \delta{\sigma} 
            +\nu \delta^2\sin\phi =0,  
\label{eQBk}
\end{equation}
\end{mathletters}
which correspond to (\ref{eQAb}) and (\ref{eQAc}).
 
     Now we can apply the formulation developed 
in the last chapter to (\ref{eQBi}), (\ref{eQBj}) and (\ref{eQBk}), 
we obtain the dissipative field equation to (\ref{eQBa}):
\begin{equation}
     \partial^2\phi(x) +\frac{H}{\nu} \sin\phi 
       -\nu\sin\phi \partial^2\delta{{\tilde \sigma}} 
       +\nu\cos\phi \partial^2\delta{{\tilde \pi}} =0,  
\label{eQBl}
\end{equation}
where $\delta{{\tilde \sigma}}$ and $\delta{{\tilde \pi}}$ 
are the similar functions as (\ref{eQAi}). 

     For $G(x-y;m)$, we take the advanced Green function
\begin{equation}
     G_{adv}(x-y;m) =\int\frac{d^4k}{(2\pi)^4} 
     \frac{e^{-ik(x-y)}}{k^2-m^2 -i\epsilon{\rm sgn}(k^0)}.  
\label{eQBm}
\end{equation}
The reality of $\delta{\tilde \sigma}$ and $\delta{{\tilde \pi}}$ 
can be checked easily by the direct calculation. 
In the next chapter, 
we will show that the advanced Green function really gives 
the dissipative solutions. 
From (\ref{eQBm}), we can read off the imaginary part of $G(k;m)$:
\begin{equation}
     i{\rm Im}{G(k;m)} =i\pi{\rm sgn}(k^0) \delta^4(k^2-m^2).  
\label{eQBn}
\end{equation}
so that $\delta{\tilde \sigma}$ and $\delta{\tilde \pi}$ become
\begin{mathletters}
\label{eQBop}
\begin{equation}
     \partial^2\delta{\tilde \sigma}(x)
     =\frac{1}{2} \int_0^\infty{dy^0} \int{d^3y} 
     \delta(x-y;m_\sigma) \partial_y^2\cos\phi(y), 
\label{eQBo}
\end{equation}
\begin{equation}
     \partial^2\delta{\tilde \pi}(x)
          =\frac{1}{2} \int_0^\infty{dy^0} \int{d^3y} 
          \delta(x-y;m_\sigma) \partial_y^2\sin\phi(y), 
\label{eQBp}
\end{equation}
\end{mathletters}
where $\delta(x-y;m)$ is the invariant delta function:
\begin{equation}
     \delta(x-y;m) =\frac{1}{i(2\pi)^3} 
                 \int{d^4k} \delta^4(k^2-m^2) {\rm sgn}(k^0) 
                 e^{-ik(x-y)}.  
\label{eQBq}
\end{equation}
%
%
\section{Solution of Dissipative Field Equation}
%
%
\subsection{Asymptotic Behavior of Solutions}
     We consider the asymptotic behavior of the solutions 
that satisfy Eq. (\ref{eQBl}) 
and show the dissipative nature of it. 

     The order parameter $\phi(x)$ is assumed to 
decrease when $t \to \infty$, 
so that it can be regarded small. 
Then we can expand Eq. (\ref{eQBl}) about $\phi$ 
and approximate to the first order (the linear approximation). 
Instead of $\phi(t,{\bf x})$, we use the Fourier component:
\begin{equation}
     \phi_{\bf k}(t) 
          =\frac{1}{(2\pi)^3} 
           \int{d^3{\bf x}} \phi(t,{\bf x}) 
           e^{-i{\bf k}{\bf x}}, 
\label{eQCa}
\end{equation}
for which Eq. (\ref{eQBl}) is approximated to be
\begin{equation}
     \ddot{\phi}_{\bf k}(t) 
     +{\bf k}^2 \phi_{\bf k}(t) 
     +\frac{1}{2} \frac{m_\pi^2}{\omega_{\bf k}}
      \int_t^\infty{ds} \sin\omega_{\bf k}(t-s) 
           \left\{ \ddot{\phi}_{\bf k}(s) 
               +{\bf k}^2 \phi_{\bf k}(s) \right\} =0,  
\label{eQCb}
\end{equation}
with $\omega_{\bf k}=\sqrt{m_\pi^2+{\bf k}^2}$ 
and $\alpha =f_\pi/\nu \sim 1.05$.\footnote{
     This slight shift from one is resulted 
     from the deformation of the chiral circle  
     because of the explicit symmetry breaking.}
It should be noted that no correlations exist among
different-momentum modes in (\ref{eQCb}) 
in the linear approximation. 

     Substituting the ansatz 
$\phi_{\bf k}(t) =e^{-a({\bf k}) t}$ into (\ref{eQCb}), 
we obtain the characteristic equation for the index $a({\bf k})$:
\begin{equation}
     a({\bf k})^4 +(2\kappa +1/2+\alpha) m_\pi^2 a({\bf k})^2 
     +\left\{ \kappa^4+(1/2+\alpha)\kappa^2 +\alpha \right\} m_\pi^4 =0,  
\label{eQCc}
\end{equation}
with $\kappa =|{\bf k}|/m_\pi$.
One of the solutions can be written as 
$a({\bf k}) =p(\kappa)+i q(\kappa)$:
\begin{equation}
     p(\kappa) =\frac{m_\pi}{2} 
             \sqrt{2\rho(\kappa) -(2\kappa^2 +1/2+\alpha)},  \quad
     q(\kappa) =\frac{m_\pi}{2} 
             \sqrt{2\rho(\kappa) +(2\kappa^2 +1/2+\alpha)},  
\label{eQCd}
\end{equation}
where $\rho(\kappa) =\sqrt{\kappa^4+(1/2+\alpha)\kappa^2+\alpha}$. 
It can be shown easily that $p(\kappa)$ takes real and positive value 
when $\kappa \geq 0$, 
so that the asymptotic behavior of $\phi_{\bf k}$ is found to be 
\begin{equation}
     \phi_{\bf k}(t) \sim e^{-p(\kappa)t} 
                          \sin[q(\kappa)t +\delta],  
          \quad (t \to \infty)
\label{eQCe}
\end{equation} 
with the constant phase. 
Summarizing the whole results, 
we obtain the asymptotic differential equation to (\ref{eQBl}):
\begin{equation}
     \ddot{\phi}_{\bf k} 
    -{\bf k}^2 \phi_{\bf k} 
    +\gamma \dot{\phi}_{\bf k} 
    +\sqrt{\alpha} \sin\phi \sim 0.   
\label{eQCf}
\end{equation}
where the dissipative coefficients are $\gamma =2p(\kappa)$. 
The momentum dependence of $p(\kappa)$ is shown in Fig.~\ref{Fig.1}.
\subsection{Numerical Results}
     The fields $\delta{\tilde \sigma}$ and $\delta{\tilde \pi}$ 
defined by (\ref{eQBo}) and (\ref{eQBp}) satisfy 
the differential equations:
\begin{equation}
     (\partial^2+m_\sigma^2) \delta{\tilde \sigma} 
          +\partial^2\cos\phi =0,  \quad
     (\partial^2+m_\pi^2) \delta{\tilde \pi} 
          +\partial^2\sin\phi =0. 
\label{eQCg}
\end{equation}
Hence the solutions of the original integro-differential 
equations are given 
as those of three differential equations, 
(\ref{eQBl}) and (\ref{eQCg}).

     In this paper, we show the numerical solutions 
for two cases: 
the uniform and the expanding solutions. 
For physical quantities, we took: 
\begin{equation}
     f_\pi =92.5{\,{\rm MeV}},  \quad  
     M     =940{\,{\rm MeV}},   \quad
     m_\pi =135{\,{\rm MeV}},   
\label{eQCh}
\end{equation}
and $m_\sigma =600{\,{\rm MeV}}$ was used for the mass of sigma meson. 
With these values, the parameters in (\ref{eQAc}) are fixed as 
$\lambda =20.0$ and $\nu =87.4{\,{\rm MeV}}$, each other.

\vskip 0.5cm
\noindent{1) {\it Uniform solution}}. 

     The solution 
that is uniform for the space-dependence 
is characterized by $\phi=\phi(t)$ 
(and correspondingly, 
$\delta{\tilde \sigma} =\delta{\tilde \sigma}(t)$ 
and $\delta{\tilde \pi} =\delta{\tilde \pi}(t)$). 
In this case, 
eqs. (\ref{eQBl}) and (\ref{eQCg}) are reduced 
to the ordinary differential equations,  
which can be easily solved. 
The numerical results are shown in Fig.~\ref{Fig.2} 
where the scaled time $\xi =m_\pi t$ has been used. 
Herein, the damping oscillation behavior 
is easily confirmed that is proved analytically 
in the last chapter. 

     The dissipating behaviors can be read off 
in the phase diagram (Fig.~\ref{Fig.3}), too, 
where each lines are the phase trajectories to these solutions; 
the spiral pattern around the origin show 
that they behave like the damping oscillator asymptotically. 

     For quantitative check, 
we consider the damping rigid oscillator: 
\begin{equation}
     {\ddot \phi} +\gamma {\dot \phi} +\sin{\phi} =0,  
\label{eQCi}
\end{equation}
with the damping coefficient $\gamma$ 
consistent with the dissipative coefficient $p(k)$ in (\ref{eQCd}):
\begin{equation}
     \gamma =2 p(0) = 0.7 m_\pi.  
\label{eQCj}
\end{equation}
The phase diagrams for (\ref{eQCi}) are given in Fig.~\ref{Fig.4}. 
The trajectories in Fig.~\ref{Fig.3} are found to behave similar 
with those in Fig.~\ref{Fig.4}, especially in the asymptotic region 
(close to the origin). 
In the non-asymptotic region (far from the origin), 
the trajectories in Fig.~\ref{Fig.3} have the modulations 
around those of the damping rigid oscillator. 
They come from the non-linear dissipating behavior 
in (\ref{eQBl}) and (\ref{eQCg}), 
which is more effective in the non-asymptotic region. 

     Through the comparison with the damping rigid oscillator, 
we can also realize the complicated behaviors 
at $(\phi=\pm\pi, \dot{\phi}=0)$ in Fig.~\ref{Fig.3}. 
They are just the turning points of the rigid oscillator, 
and the trajectories around them are changed in chaotic manner 
under the small perturbation. 

\vskip 0.5cm
\noindent {2) {\it Expanding solution}}. 

     With putting $\phi =\phi(\tau =\sqrt{t^2-x^2})$ 
in (\ref{eQBl}) and (\ref{eQCg}), 
we get the expanding solution in $x$-direction 
(uniform in other directions). 
Originally, this type of solution is given 
by Blaizot and Kryzwicki \cite{BK}
with no dissipative effect. 
 
     The numerical solution is given as the rigid line 
in Fig.~\ref{Fig.5}, 
where the scaled local time $\xi =m_\pi \tau$ was used. 
In this figure, 
we can find that the expanding solution is dumped faster 
than the uniform solution 
(shown as the dotted line in Fig.~\ref{Fig.4}).  
To realize the expansion effect, 
we study the differential equation 
that Blaizot and Kryzwicki solved. 
In our notation, it becomes
\begin{equation}
     \phi''+\frac{1}{\xi} \phi' +\sin\phi =0,  
\label{eQCk}
\end{equation}
where the differentiations are for $\xi=m_\pi\tau$. 
The second term in (\ref{eQCk}) is proportional to 
the time-derivative of $\phi$, 
and has the effect of dissipation 
with the time-dependent dissipative coefficient $1/\xi$. 
(This effect is not a real dissipation, 
but a smearing of $\phi$ brought by the volume expansion.)
In the present case, 
this smearing effect has an additional effect with 
the real dissipative effect (represented by the second term 
in (\ref{eQCi}) asymptotically),
and it causes the faster damping in the expanding solution. 
The smearing effect is found to be more effective 
in the non-asymptotic region, 
because the effective dissipating coefficient 
is inversely proportional with the local time $\tau$.
%
%
\section{Summary and Discussions}
%
%
     We have formulated the dissipative field theory 
with applying the Caldeira-Leggett method. 
Explicit calculations have been done for the linear sigma model 
and the resultant field equations have been shown 
to have the dissipative properties 
with both the analytic and numerical ways. 

     As a phenomenological application, 
we discuss about the disoriented chiral condensate 
that is expected to appear 
after the high-energy hadron collision. 
In the standard picture, 
the chiral symmetry is considered to be broken spontaneously 
at zero temperature 
and its order parameters take the expectation values:  
${\langle \sigma \rangle} \neq 0$ and ${\langle \pi \rangle}=0$. 
The DCC is also in the broken phase, 
but is defined to be the state 
where the order parameters take different values: 
${\langle \sigma \rangle} \neq 0$ and 
${\langle \pi \rangle} \neq 0$. 
Because of the explicit chiral-symmetry breaking, 
the DCC state has higher energy, 
and it should be observed as a metastable state. 

     As written in the introduction, 
there exist many controversies 
about the formation process of the DCC state 
after rebreaking of the chiral symmetry, 
but we concentrate on the decay process of it 
in the remaining of this paper. 

     We consider the neutral DCC state 
with 
${\langle \pi^0 \rangle} \neq 0$ and 
${\langle \pi^\pm \rangle} =0$.\footnote{
For the charged DCC state, 
photon degree of freedom is absolutely important 
and we have to extend our equations 
to include the dissipative effect 
by photon radiation.}
The main decay process of this state 
should be the $\pi^0$-radiation, 
so that we can apply the above-developed formula 
with regarding $\pi$ in (\ref{eQCf}) as $\pi^0$. 
The life-time $\tau_L$ can be estimated 
with the dissipative constant $p(0)$ in (\ref{eQCd}):  
\begin{equation}
     \tau_L =1/p(0) \sim 3 m_\pi^{-1}.     
\label{eQCl}
\end{equation}
If we take the quenching scenario for the DCC formation, 
the formation time $\tau_R$ has been estimated 
to be \cite{RW}
\begin{equation}
    \tau_R \sim \sqrt{2} m_\sigma^{-1} \sim 0.3 m_\pi^{-1}.  
\label{eQCm}
\end{equation}
It tells us that the life-time $\tau_L$ 
in (\ref{eQCl}) is ten times longer 
than the formation time, 
so that the neutral DCC state will be enough metastable. 

     In this paper, we considered the case that the order parameters 
move only on the chiral circle 
${\langle \sigma \rangle}^2+{\langle \pi \rangle}^2 =f_\pi^2$, 
so that we could not consider the DCC formation process. 
The extension beyond the chiral circle will be given elsewhere. 
\newpage
%
%
%
%

%
%
%
%
\begin{figure}
\caption{The momentum dependence of the dissipative coefficient 
$p(\kappa)$ given by (\protect\ref{eQCd}). 
$\kappa$ is the pion-mass scaled momentum $\kappa =|{\bf k}|/m_\pi$. 
The $p(\kappa)$ is normalized at $\kappa=0$ with the value 
$p(0)=0.35 m_\pi$.}
\label{Fig.1} 
\end{figure}

\begin{figure}
\caption{A series of solutions of 
the integro-differential equation (\protect\ref{eQBl})
for the space-independent uniform case $\phi(x) \equiv \phi(t)$. 
The $\xi$ is the pion-mass scaled time:
$\xi =m_\pi t$ and $\phi(t)$ is a chiral angle 
of the order parameter. 
The initial conditions are given at $\xi=0$ 
with $\frac{d\phi}{d\xi} =0$.}
\label{Fig.2} 
\end{figure}

\begin{figure}
\caption{The trajectories of the integro-differential equation 
(\protect\ref{eQBl}) for the space-independent uniform case 
$\phi(x) \equiv \phi(t)$. 
The $\phi$ and $\frac{d\phi}{d\xi}$ are the chiral angle 
and the corresponding velocity with the scaled time  
$\xi =m_\pi t$. 
The initial conditions of each trajectories are chosen 
as they behave asymptotically (\protect\ref{eQCe})
with $\delta =\frac{\pi n}{5}$ ($n=0,\pm1,\cdots,\pm 5$).}
\label{Fig.3} 
\end{figure}

\begin{figure}
\caption{The trajectories of the dumped rigid oscillator 
(\protect\ref{eQCi}).
The dissipative coefficient is chosen to be consistent 
with the asymptotic value
in the integro-differential equation (\protect\ref{eQBl}): 
$\gamma =2 p(0) = 0.7 m_\pi$.
The $\phi$ and $\frac{d\phi}{d\xi}$ are the chiral angle 
and the corresponding velocity with the scaled time  
$\xi =m_\pi t$. 
The initial conditions of each trajectories are chosen 
as in \protect\ref{Fig.3}. }
\label{Fig.4} 
\end{figure}

\begin{figure}
\caption{One-dimensional expanding (scaling) solutions 
for the chiral angle $\phi(\xi)$ 
where $\xi$ is a pion-mass scaled local time: 
$\xi =m_\pi \tau =m_\pi \protect\sqrt{t^2-x^2}$. 
	The rigid line is for the dissipative case:  
the solution of the integro-differential equation 
(\protect\ref{eQBl}), 
and the dashed line is for the nondissipative case: 
the original Blaizot-Kryzwicki solution of (\protect\ref{eQCk}). }
\label{Fig.5} 
\end{figure}
\end{document}